\documentclass[twocolumn,twoside,slac_two]{revtex4}
\usepackage{graphicx}
\usepackage{fancyhdr}
\def\Dzkpi{D^0\to K^-\pi^+}
\def\Dzkpipiz{D^0\to K^-\pi^+\pi^0}
\def\Dzkpipipi{D^0\to K^-\pi^+\pi^+\pi^-}
\def\Dpkpipi{D^+\to K^-\pi^+\pi^+}
\def\Dpkpipipiz{D^+\to K^-\pi^+\pi^+\pi^0}
\def\Dpkspi{D^+\to K^0_S\pi^+}
\def\Dpkspipiz{D^+\to K^0_S\pi^+\pi^0}
\def\Dpkspipipi{D^+\to K^0_S\pi^+\pi^+\pi^-}
\def\Dpkkpi{D^+\to K^+K^-\pi^+}

\begin{document}

\title{Determination of Charm Hadronic Branching Ratios and New Modes}

%

\author{A. Ryd}
\affiliation{Cornell University, Newman Laboratory, Ithaca NY 14853, USA}

\begin{abstract}
Recent results from CLEO-c, BABAR,
and Belle on measurements of absolute branching fractions
of $D$ and $D_s$ mesons are reviewed. 
\end{abstract}

\maketitle

\thispagestyle{fancy}


\section{Introduction}

 Precise measurements of the absolute branching fractions
for $D$ and $D_s$ meson decays are important as they serve
to normalize most $B$ and $B_s$ decays as well as many
charm decays.  Recent measurements from
CLEO-c, BABAR, and Belle for the measurements of the
absolute hadronic branching fractions of $D$ and $D_s$
mesons are presented here. 

Results from the CLEO-c experiment at the Cornell 
Electron Positron Storage Ring
based on 281 pb$^{-1}$ recorded at the $\psi(3770)$ are
presented here for studies of $D^0$ and $D^+$ decays. 
In addition, CLEO-c has analyzed 195 pb$^{-1}$
of $e^+e^-$ annihilation data near $E_{\rm cm}=4170$ MeV for
studies of $D_s$ decays.
These samples
provide very clean environments for studying decays of 
$D$ and $D_s$ mesons. 
The $\psi(3770)$
produced in the $e^+e^-$ annihilation decays to pairs of $D$ mesons,
either $D^+D^-$ or $D^0\bar D^0$. In particular, the
produced $D$ mesons can not be accompanied
by any additional pions. At $E_{\rm cm}=4170$ MeV $D_s$ 
mesons are primarily produced as $D_s^{+}D_s^{*-}$ and
$D_s^{*+}D_s^{-}$ pairs. 

The results from BABAR and Belle use their large 
samples of $e^+e^-$ data collected by these experiments. The
different analyses presented here use integrated luminosities
up to 0.55 ab$^{-1}$. For example, Belle has used 0.55 ab$^{-1}$
to study $D_s^+\to K^+K^-\pi^+$ in exclusive 
production of $e^+e^-\to D_s^*D_{s1}$. BABAR has studied 
$D_s\to \phi\pi$ using a sample of $B\to D^{(*)}D_{s(J)}^{(*)}$ decays.
These examples illustrate that charm produced both in the
continuum and in $B$ meson decays are useful for studies of
charm at the $B$-factories.

First I will discuss the determination of the absolute
$D^0$ and $D^+$ branching fractions. New results from 
CLEO-c and BABAR are discussed here. Then results for
$D_s$ branching fractions from CLEO-c, Belle, and BABAR
are presented. Last a few inclusive and rare hadronic 
decay modes are discussed.

\section{Absolute $D$ hadronic branching fractions at CLEO-c}

This analysis makes use of a 'double tag' technique
initially used by Mark III~\cite{markiii}.
In this technique the yields of single tags, where
one $D$ meson is reconstructed per event, and
double tags, where both $D$ mesons are reconstructed,
are determined.
The number of single tags,
separately for $D$ and $\bar D$ decays, are given by
$N_i=\epsilon_i{\cal B}_i N_{D\bar D}$ and 
${\bar N}_j=\bar \epsilon_j{\cal B}_j N_{D\bar D}$
where $\epsilon_i$ and ${\cal B}_i$ are the efficiency
and branching fraction for mode $i$. Similarly,
the number of double tags reconstructed are given
by $N_{ij}=\epsilon_{ij}{\cal B}_i{\cal B}_j N_{D\bar D}$
where $i$ and $j$ label the $D$ and $\bar D$ mode used
to reconstruct the event and $\epsilon_{ij}$ is the
efficiency for reconstructing the final state.
Combining the equations above and solving for $N_{D\bar D}$
gives the number of produced $D\bar D$ events 
as 
$$
N_{D\bar D}={{N_i}{\bar N_j}\over N_{ij}}{\epsilon_{ij}\over \epsilon_i\bar\epsilon_j}
$$
and the branching fractions 
$$
{\cal B}_i={N_{ij}\over N_j}{{\epsilon_j}\over \epsilon_{ij}}.
$$
In this analysis CLEO-c determine all the
single tag and double tag yields in data, determine the efficiencies
from Monte Carlo simulations of the detector
response, and extract the branching
fractions and $D\bar D$ yields from a combined fit to all 
measured data yields.

\begin{figure}[tb]
\begin{center}
\includegraphics[width=\linewidth]{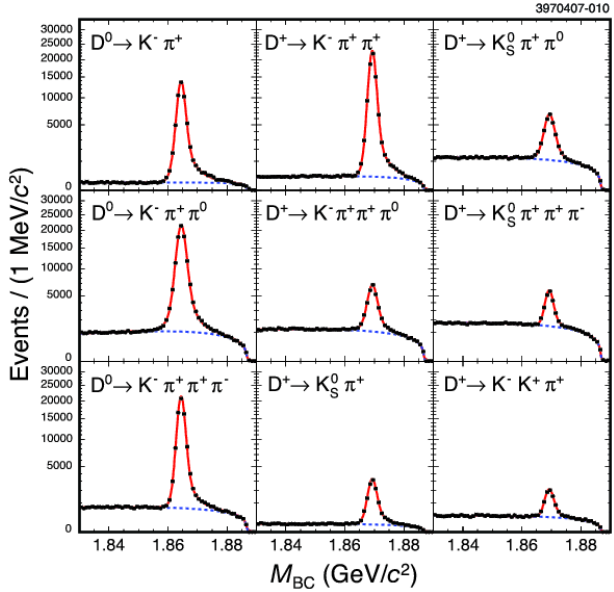}
\caption{The fits for the single tag yields. The background is 
described by the ARGUS threshold function and the signal shape
includes the effects of beam energy spread, momentum resolution,
initial state radiation, and the $\psi(3770)$ lineshape. }
\label{fig:dhad_st}
\end{center}
\end{figure}

\begin{figure}[tb]
\begin{center}
\includegraphics[width=0.49\linewidth]{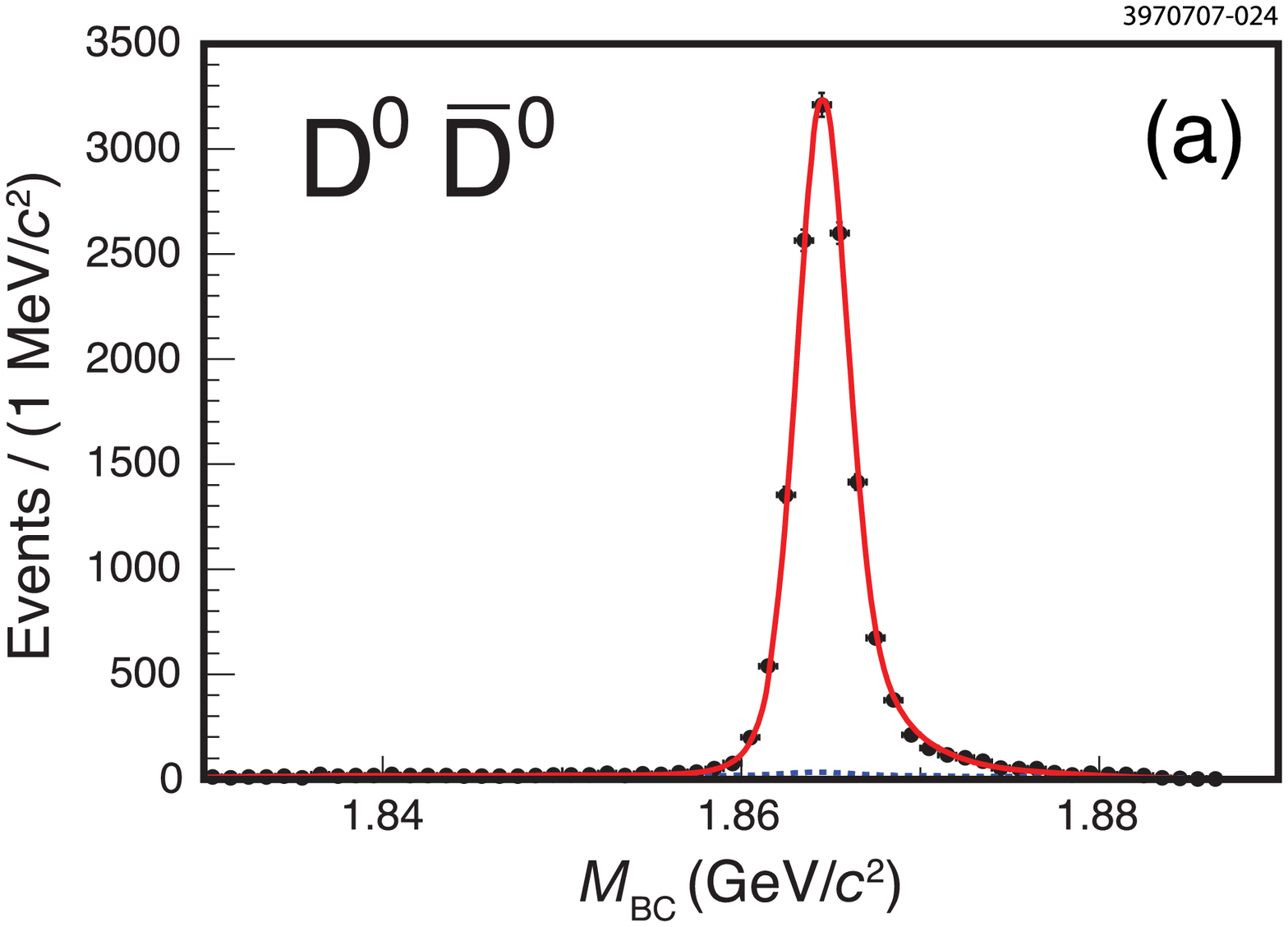}
\includegraphics[width=0.49\linewidth]{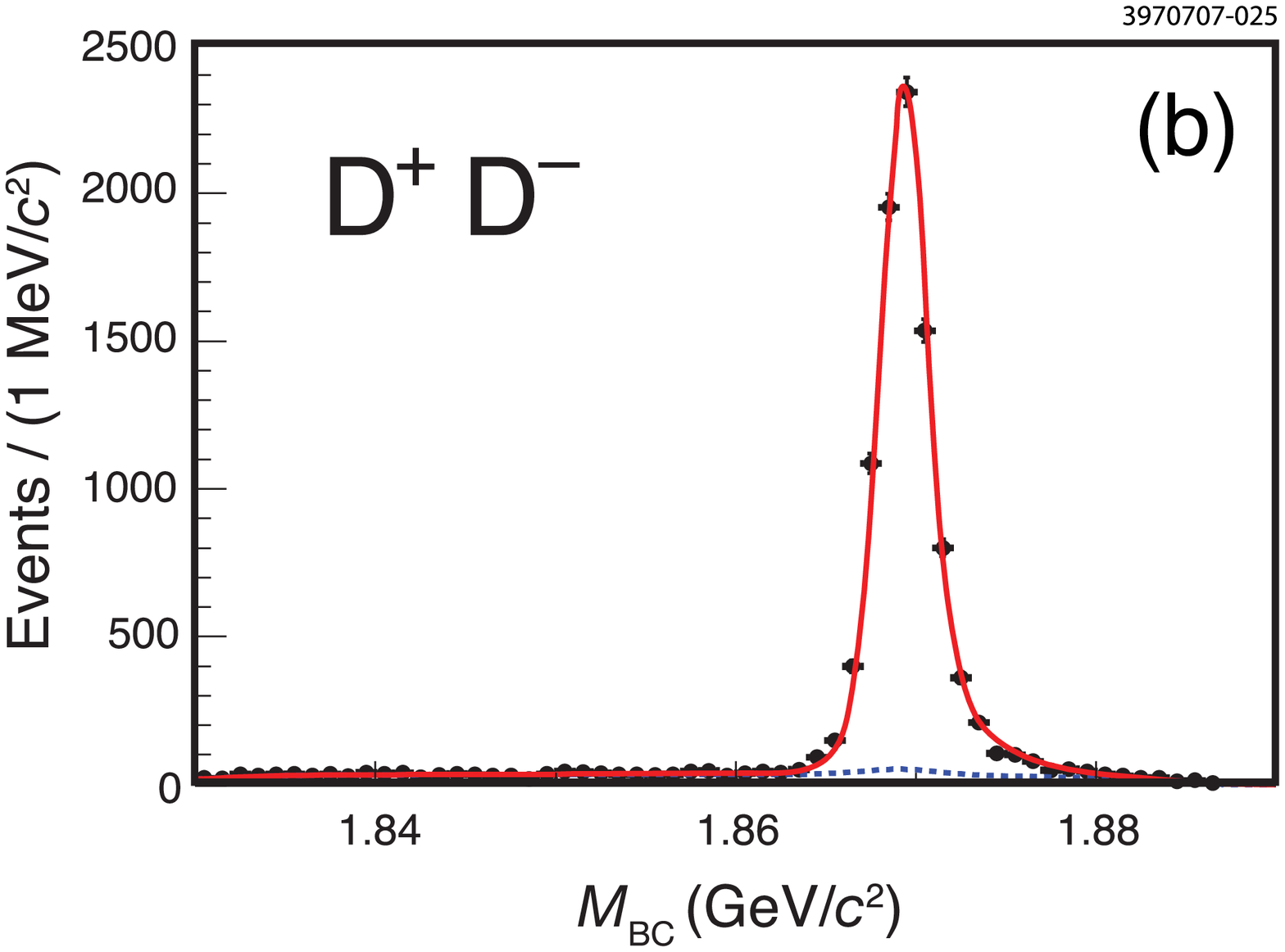}
\caption{The fit for the double tag yields combined over
all modes for charged and neutral modes separately.}
\label{fig:dhad_dt}
\end{center}
\end{figure}

This analysis uses three $D^0$ decays 
($D^0\to K^-\pi^+$, $D^0\to K^-\pi^+\pi^0$, and $D^0\to K^-\pi^+\pi^-\pi^+$) 
and six $D^+$ modes
($D^+\to K^-\pi^+\pi^+$, $D^+\to K^-\pi^+\pi^+\pi^0$, $D^+\to K^0_S\pi^+$, $D^+\to K^0_S\pi^+\pi^0$,
 $D^+\to K^0_S\pi^+\pi^-\pi^+$, and $D^+\to K^-K^+\pi^+$).
The single tag yields are shown in Fig.~\ref{fig:dhad_st}.
The combined double tag
yields are shown in Fig.~\ref{fig:dhad_dt}
for charged and neutral $D$ modes separately.
The scale of the statistical errors on the branching fractions
are set by the number of double tags and  
precisions of 
$\approx 0.8\%$ and $\approx 1.0\%$ are obtained for the neutral and charged modes
respectively.
The branching fractions obtained are summarized in Table~\ref{tab:dhadresults}.

CLEO-c has presented updated results for these 
branching fractions\cite{cleoc_charm07} since these 
results were presented. The new results, including
${\cal B}(D^0\to K^-\pi^+)=(3.891\pm 0.035\pm 0.059\pm 0.035)\%$, are 
consistent with the preliminary results presented here. The last error
is the uncertainty due to final state radiation.

\begin{table}[bt]
\caption{Preliminary branching fractions from CLEO-c.  
Uncertainties are statistical and systematic,
respectively. 
}
\label{tab:dhadresults}
\begin{center}
\begin{tabular}{lcc}
\hline\hline
Mode & Fitted Value (\%)& PDG (\%)  \\
\hline
${\cal B}(\Dzkpi)$        & $3.87\pm 0.04\pm 0.08$        & $3.81\pm0.09$ \\
${\cal B}(\Dzkpipiz)$     & $14.6\pm 0.1\pm 0.4$          & $13.2\pm1.0$ \\
${\cal B}(\Dzkpipipi)$    & $8.3\pm 0.1\pm 0.2$           & $7.48\pm0.30$ \\
\hline
${\cal B}(\Dpkpipi)$      & $9.2\pm 0.1\pm 0.2$           & $9.2\pm0.6$ \\
${\cal B}(\Dpkpipipiz)$   & $6.0\pm 0.1\pm 0.2$           & $6.5\pm1.1$ \\
${\cal B}(\Dpkspi)$       & $1.55\pm 0.02\pm 0.05$        & $1.42\pm0.09$ \\
${\cal B}(\Dpkspipiz)$    & $7.2\pm 0.1\pm 0.3$           & $5.4\pm1.5$ \\
${\cal B}(\Dpkspipipi)$   & $3.13\pm 0.05\pm 0.14$           & $3.6\pm0.5$ \\
${\cal B}(\Dpkkpi)$       & $0.93\pm 0.02\pm 0.03$        & $0.89\pm0.08$ \\
\hline\hline
\end{tabular}
\end{center}
\end{table}

\section{Measurement of ${\cal B}(D^0\to K^-\pi^+)$ at BABAR}

\begin{figure}[tb]
\begin{center}
\includegraphics[width=0.9\linewidth]{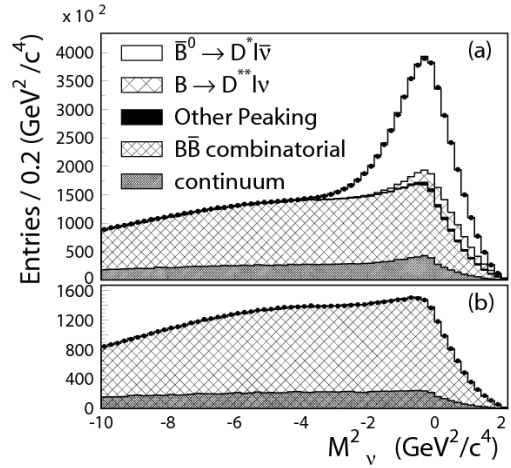}
\caption{The distribution of the missing mass squared, $M^2_{\nu}$,
for (a) right sign events and (b) wrong sign events. The wrong sign
events show that the simulation of the background shape is good.
(From Ref.~\cite{babar_dtokpi}.) }
\label{fig:babar_m2nu}
\end{center}
\end{figure}

BABAR has used a sample of 210 fb$^{-1}$ of $e^+e^-$ data
collected at the $\Upsilon(4S)$ resonance to study the 
decay $D^0\to K^-\pi^+$ decay~\cite{babar_dtokpi}. They use semileptonic 
$B$ decays, $\bar B^0\to D^{*+}\ell^-\bar\nu$ followed
by $D^{*+}\to D^0\pi^+$, where they use the lepton
in the $B$ decay and the slow pion from the $D^*$ to 
tag the signal. As the energy release in the $D^*$ decay
is very small the reconstructed slow pion momentum can be used
to estimate the four-momentum of the $D^*$ --- the slow pion
and the $D^*$ have approximately the same velocity.
BABAR extracts the number of $\bar B^0\to D^{*+}\ell^-\bar\nu$
decays using the missing mass squared, $M^2_{\nu}$, against the
$D^*$ and the lepton. The $M^2_{\nu}$ distribution 
is shown in Fig.~\ref{fig:babar_m2nu}.
A clear signal is observed for $M^2_{\nu}>-2.0$ GeV$^2$. 
However, there are substantial backgrounds that need
to be subtracted due to combinatorial backgrounds
in $B\bar B$ events and continuum production. Table~\ref{tab:babar_bkgd}
summarizes the event yields for the inclusive 
$\bar B^0\to D^{*+}\ell^-\bar\nu$ reconstruction in the column 
labeled 'Inclusive'.
BABAR finds $2,170,640\pm 3,040$ $\bar B^0\to D^{*+}\ell^-\bar\nu$
decays followed by $D^{*+}\to D^0\pi^+$.

\begin{table}[bt]
\caption{Event yields for the inclusive $\bar B^0\to D^{*+}\ell^-\bar\nu$
reconstruction and the exclusive analysis where the $D^0\to K^-\pi^+$
final state is reconstructed in the BABAR analysis to determine the 
branching fraction for $D^0\to K^-\pi^+$ decay.
}
\label{tab:babar_bkgd}
\begin{center}
\begin{tabular}{lcc}
\hline\hline
Source & Inclusive & Exclusive  \\
\hline
Data                       & $4,412,390\pm 2100$        & $47,270\pm 220$ \\
Continuum                  & $460,030\pm 2090$          & $3,090\pm 170$ \\
Combinatorial $B\bar B$    & $1,781,720\pm 680$         & $8,190\pm 50$ \\
Peaking                    &                            & $1,630\pm 80$ \\
Cabibbo suppressed         &                            & $550\pm 10$ \\
\hline
Signal                     & $2,170,640\pm 3,040$       & $33,810\pm 290$ \\
\hline\hline
\end{tabular}
\end{center}
\end{table}

\begin{figure}[tb]
\begin{center}
\includegraphics[width=0.9\linewidth]{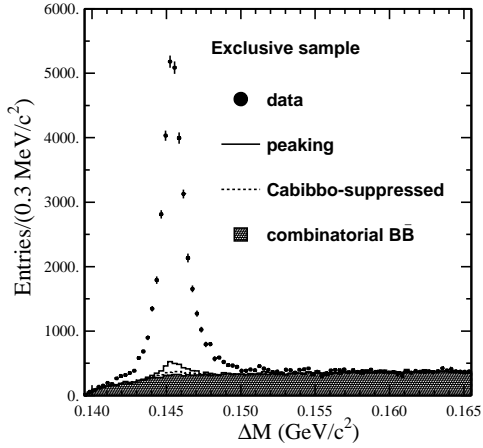}
\caption{The $\Delta M$ distribution for the reconstructed $D^0\to K^-\pi^+$
candidates in events with a $\bar B^0\to D^{*+}\ell^-\bar\nu$ tag.
(From Ref.~\cite{babar_dtokpi}.)}
\label{fig:babar_dm}
\end{center}
\end{figure}

The next step in this analysis is to use this sample of events and
reconstruct the $D^0\to K^-\pi^+$ decay. To extract a clean signal
BABAR studies the mass difference $\Delta M\equiv m_{K\pi\pi_s}-m_{K\pi}$
where $\pi_s$ indicate the slow pion from the $D^*$ decay. The 
mass difference is shown in Fig.~\ref{fig:babar_dm}. The yields
for this 'Exclusive' analysis are given in Table~\ref{tab:babar_bkgd}.
Using simulated events BABAR determine an efficiency of $(39.96\pm0.09)\%$
for reconstructing the $D^0\to K^-\pi^+$ final state. Combining this
with the data yields given above BABAR determines
$$
{\cal B}(D^0\to K^-\pi^+)=(4.007\pm0.037\pm0.070)\%.
$$
This is slightly larger than the branching fraction CLEO-c obtained,
but within errors they are consistent.

\section{Absolute branching fractions for hadronic $D_s$ decays at CLEO-c}

This analysis uses a sample of 195 pb$^{-1}$ of data recorded at
a center-of-mas energy of 4170 MeV. At this energy $D_s$ mesons
are produced, predominantly, as $D_s^+D_s^{*-}$ or $D_s^-D_s^{*+}$
pairs. CLEO-c uses the same tagging technique as for the hadronic
$D$ branching fractions; they reconstruct samples of single tags and
double tags and use this to extract the branching fractions. 

\begin{figure}[tb]
\begin{center}
\includegraphics[width=0.49\linewidth]{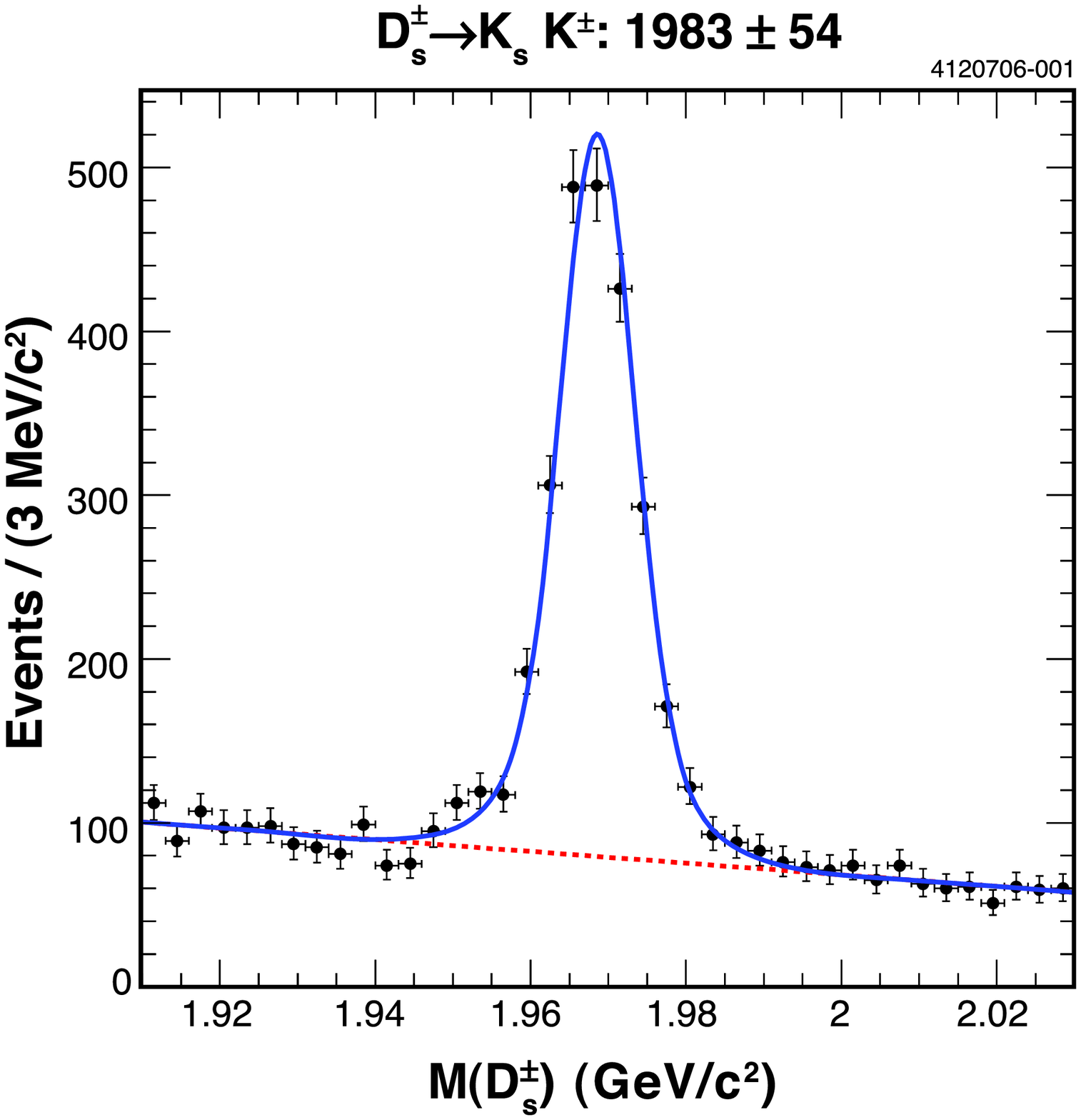}
\includegraphics[width=0.49\linewidth]{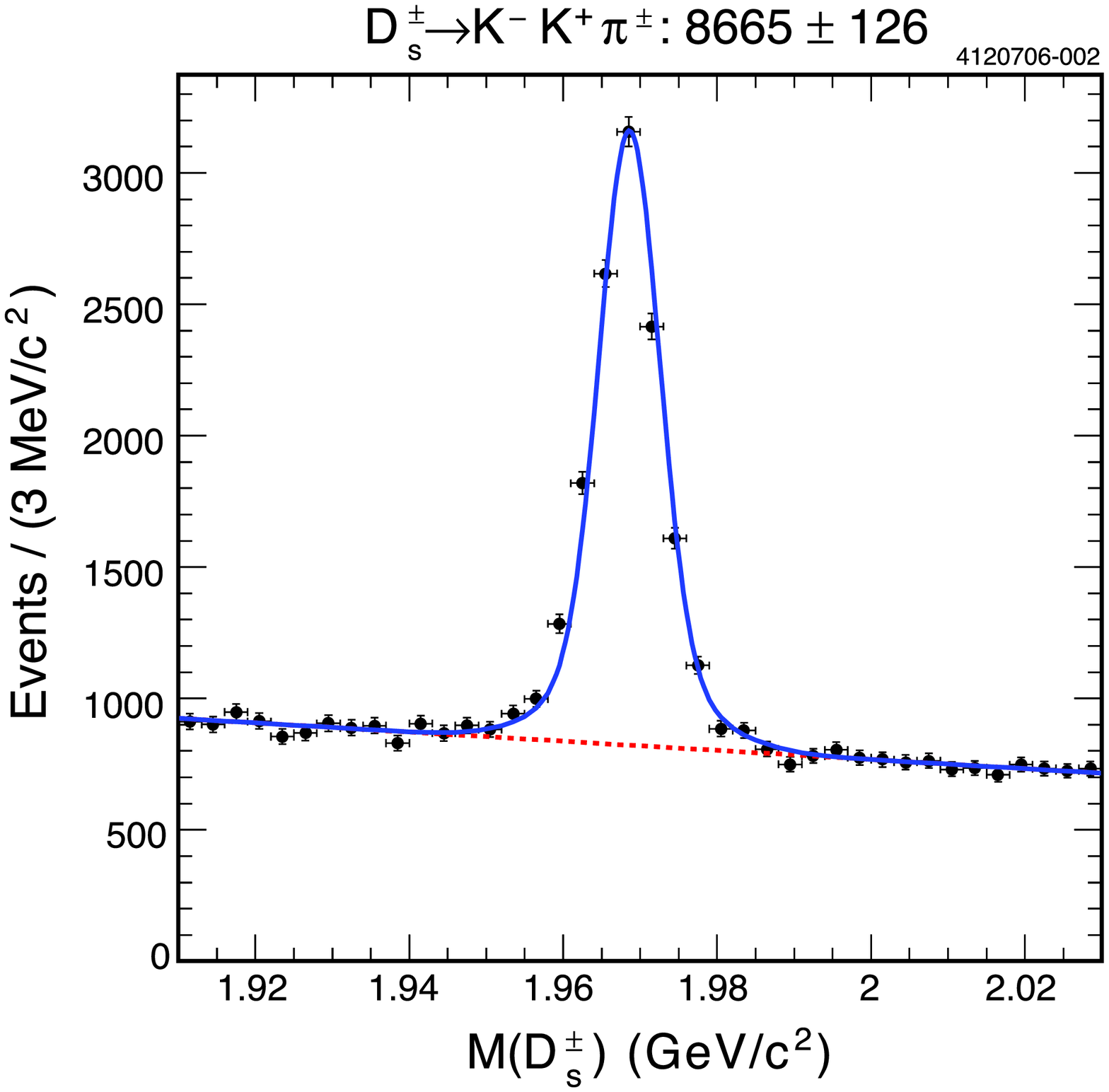}
\includegraphics[width=0.49\linewidth]{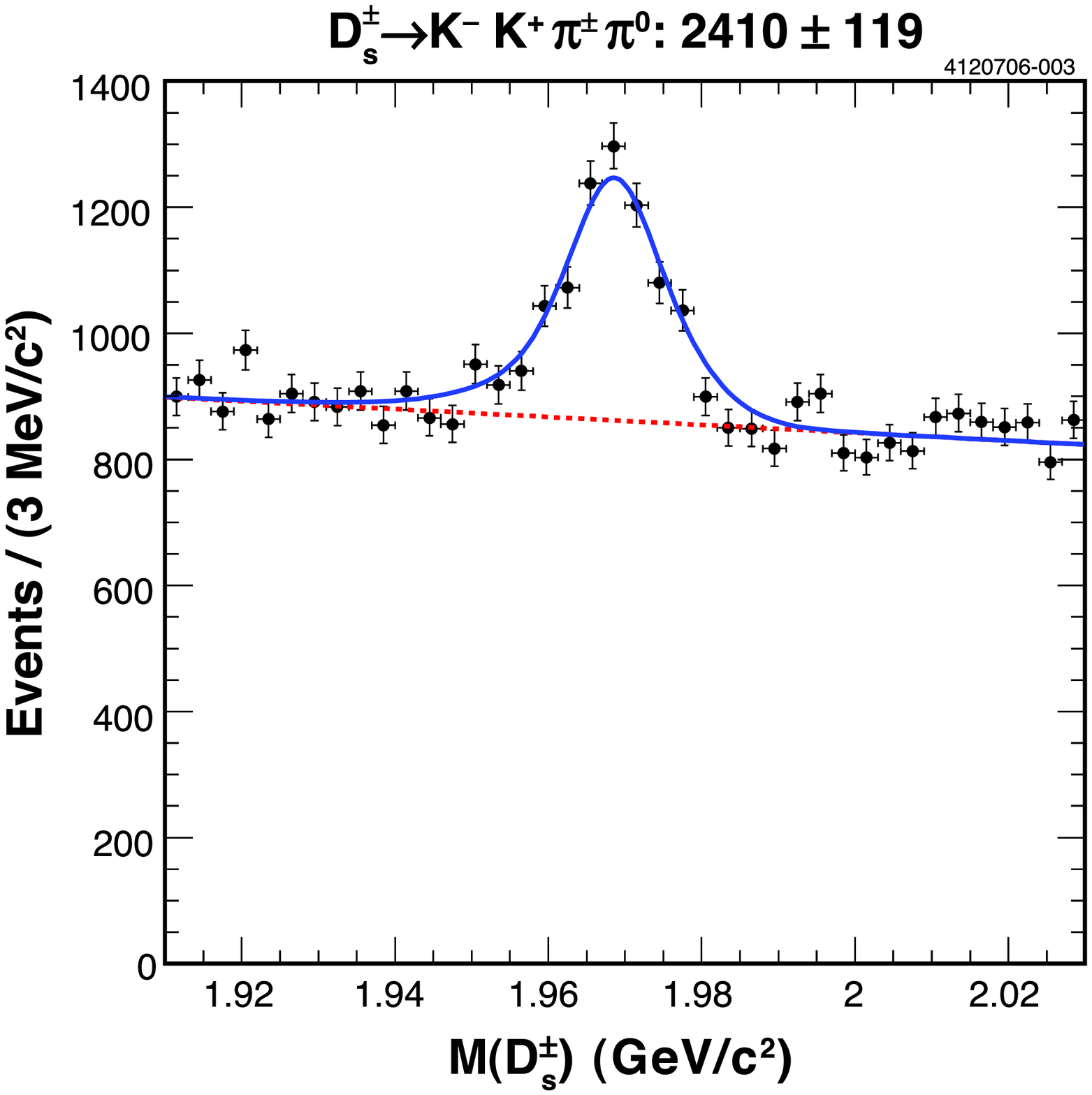}
\includegraphics[width=0.49\linewidth]{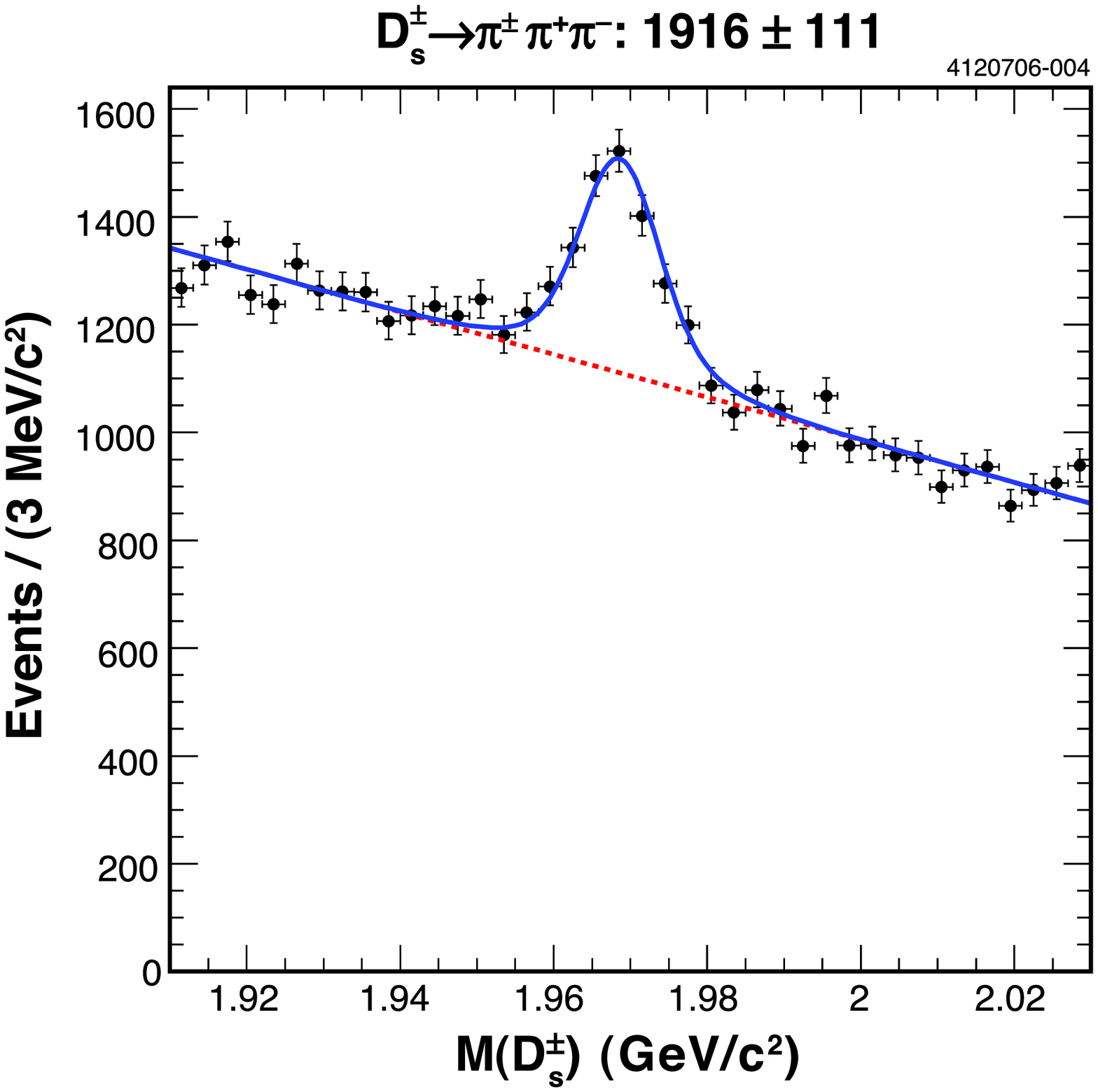}
\includegraphics[width=0.49\linewidth]{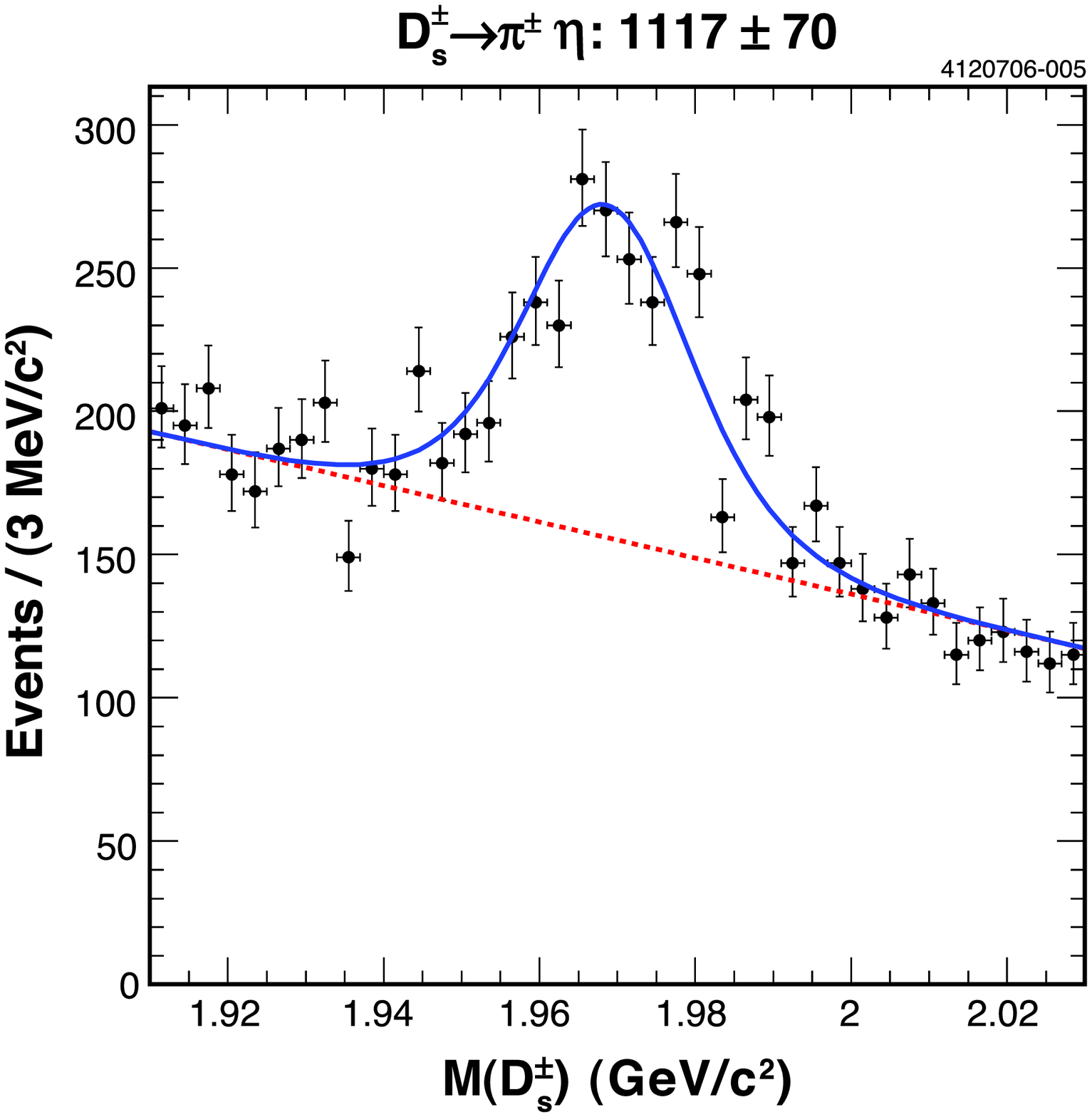}
\includegraphics[width=0.49\linewidth]{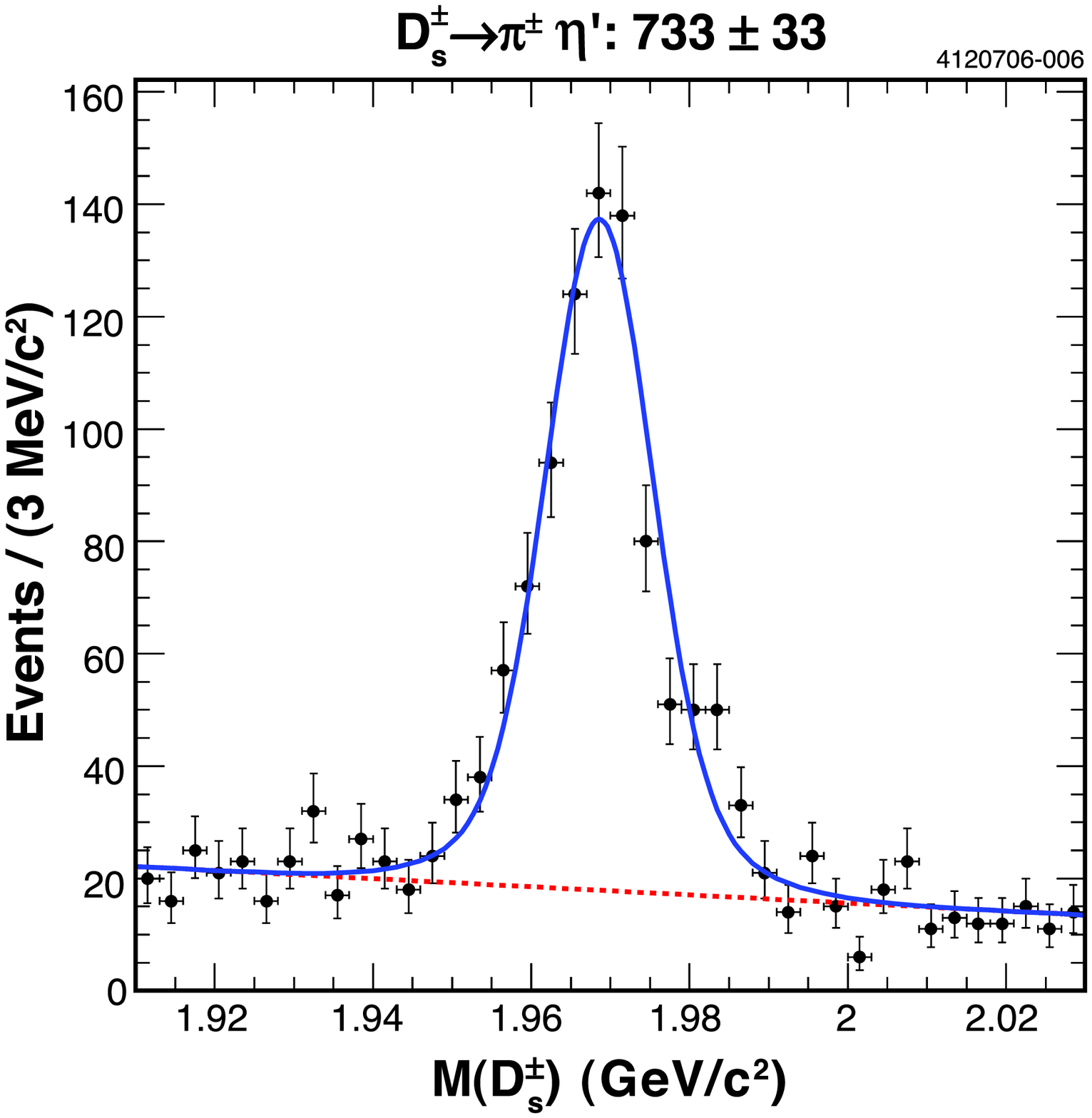}
\caption{Single tag yields for $D_s$ modes used in the CLEO-c 
analysis.}
\label{fig:cleoc_ds_st}
\end{center}
\end{figure}

\begin{figure}[tb]
\begin{center}
\includegraphics[width=0.7\linewidth]{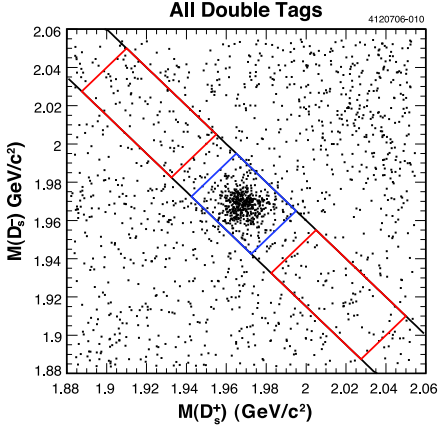}
\caption{Double tag yields for $D_s$ modes used in the CLEO-c 
analysis.}
\label{fig:cleoc_ds_dt}
\end{center}
\end{figure}

CLEO-c studies six $D_s$ final states ($D^+_s\to K^0_S K^+$,
$D^+_s\to K^+K^-\pi^+$,
$D^+_s\to K^+K^-\pi^+\pi^0$,
$D^+_s\to \pi^+\pi^-\pi^+$,
$D^+_s\to \eta\pi^+$, and
$D^+_s\to \eta'\pi^+$). The single tag event yields are shown
in Fig.~\ref{fig:cleoc_ds_st}. The double tag yields are extracted
by a cut-and-count procedure in the plot of the invariant mass
of the $D^+_s$ vs. $D^-_s$. This plot is shown in Fig.~\ref{fig:cleoc_ds_dt}.
Backgrounds are subtracted from the sidebands indicated in
the plot and a total of 471 double tag events are found. 

\begin{table}[bt]
\small{
\caption{Preliminary branching fractions for $D_s$ decays determined in the 
CLEO-c analysis.
 }
\label{tab:cleoc_ds_brfr}
\begin{center}
\begin{tabular}{lcc}
\hline\hline
Mode & Branching Fraction (\%) \\ \hline
${\cal B}(D^+_s\to K^0_S K^+)$        & $1.50\pm0.09\pm0.05$        \\
${\cal B}(D^+_s\to K^+K^-\pi^+)$      & $5.57\pm0.30\pm0.19$         \\
${\cal B}(D^+_s\to K^+K^-\pi^+\pi^0)$ & $5.62\pm0.33\pm0.51$         \\
${\cal B}(D^+_s\to \pi^+\pi^-\pi^+)$  & $1.12\pm0.08\pm0.05$         \\
${\cal B}(D^+_s\to \eta\pi^+)$        & $1.47\pm0.12\pm0.14$         \\
${\cal B}(D^+_s\to \eta'\pi^+)$       & $4.02\pm0.27\pm0.30$       \\
\hline\hline
\end{tabular}
\end{center}
}
\end{table}

From these yields CLEO-c determines the branching fractions
listed in Table~\ref{tab:cleoc_ds_brfr}.
CLEO-c is not quoting branching fractions for $D_s^+\to \phi\pi^+$
as the $\phi$ signal is not well defined. In particular, 
the $\phi$ resonance
interferes with the $f_0$ resonance. CLEO-c reports 
preliminary results for partial
branching fractions for $D_s^+\to K^+K^-\pi^+$ in restricted
invariant mass ranges of $m_{KK}$ near the $\phi$ resonance.
In particular, for a 10 MeV cut around the $\phi$ mass the
partial branching fraction of $(1.98\pm0.12\pm0.09)\%$ is
found while for a 20 MeV cut the corresponding branching
fraction is $(2.25\pm0.13\pm0.12)\%$.

Since these results were presented CLEO-c has updated 
this analysis to include 298 pb$^{-1}$ of
data recorded at the $E_{\rm cm}=4170$ MeV~\cite{cleoc_charm07}. 
In addition to the
six mode used in the analysis described above CLEO-c also
uses $D_s^+\to K^+\pi^+\pi^-$ and $D^+_s\to K^0_SK^-\pi^+\pi^+$. 
Among the updated
results is the branching fraction 
${\cal B}(D_s^+\to K^+K^-\pi^+=(5.67\pm0.24\pm0.18)\%$, in
good agreement with the preliminary result presented above.

\section{Belle study of $D_s^+\to K^+K^-\pi^+$}

Using 0.55 ab$^{-1}$ of $e^+e^-$ data recorded with
the Belle detector at KEKB the Belle collaboration
has studied the process $e^+e^-\to D_s^{*+}D^-_{s1}$
followed by $D^-_{s1}\to D^{*0}K^-$ and 
$D_s^{*+}\to D_s^+\gamma$\cite{belle_dskkpi}.
The final state is reconstructed in two ways; either
by partially reconstructing the $D_{s1}$ or the $D_s^*$.

Belle obtains the branching fraction ${\cal B}(D^+_s\to K^+K^-\pi^+)=
(4.0\pm0.4\pm0.4)\%$. This is somewhat lower than the 
CLEO-c result presented in the previous section.

\section{BABAR studies of $D_s\to \phi\pi$}

\begin{figure}[tb]
\begin{center}
\begin{minipage}{\linewidth }
\includegraphics[width=0.49\linewidth]{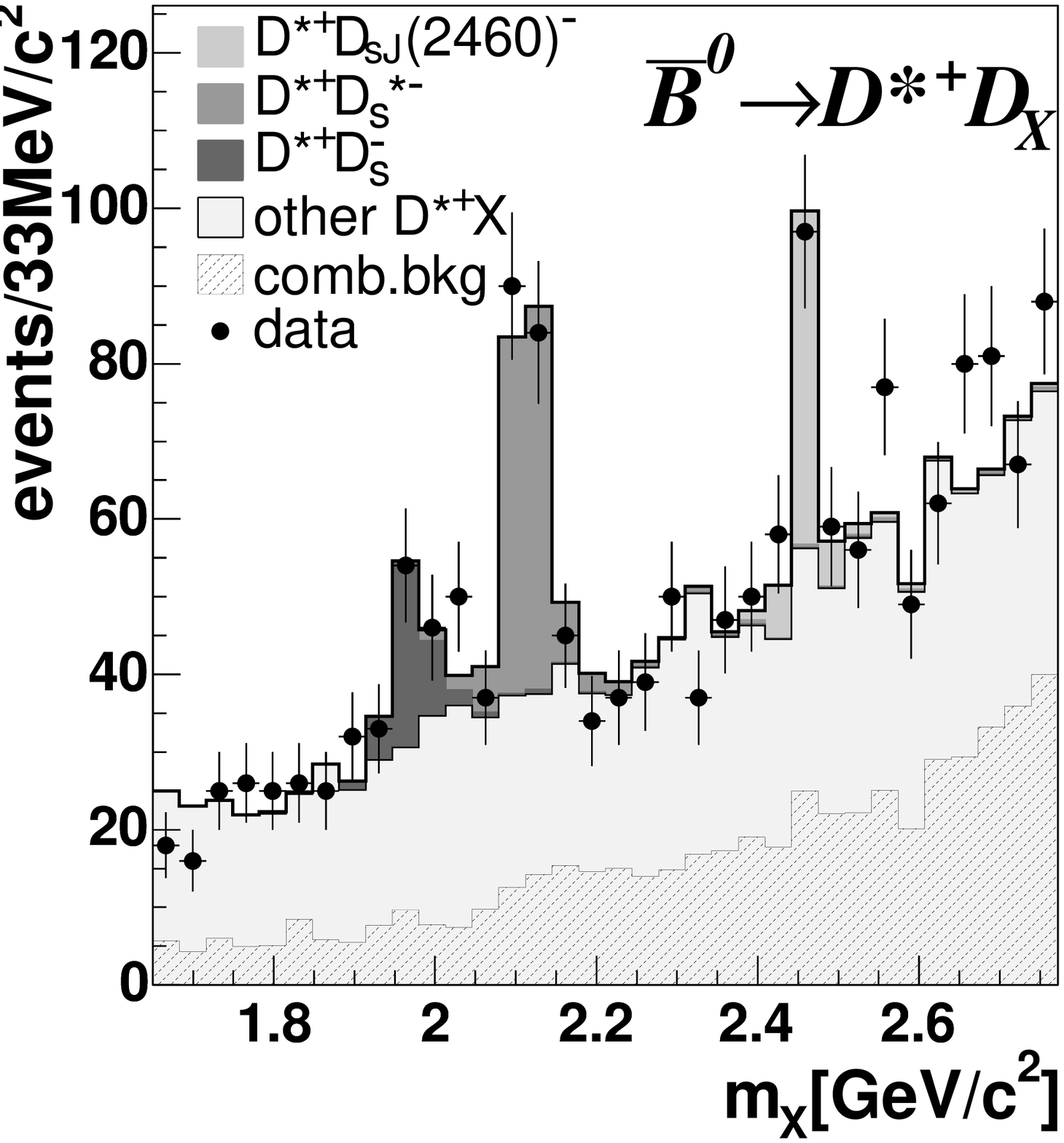}
\includegraphics[width=0.49\linewidth]{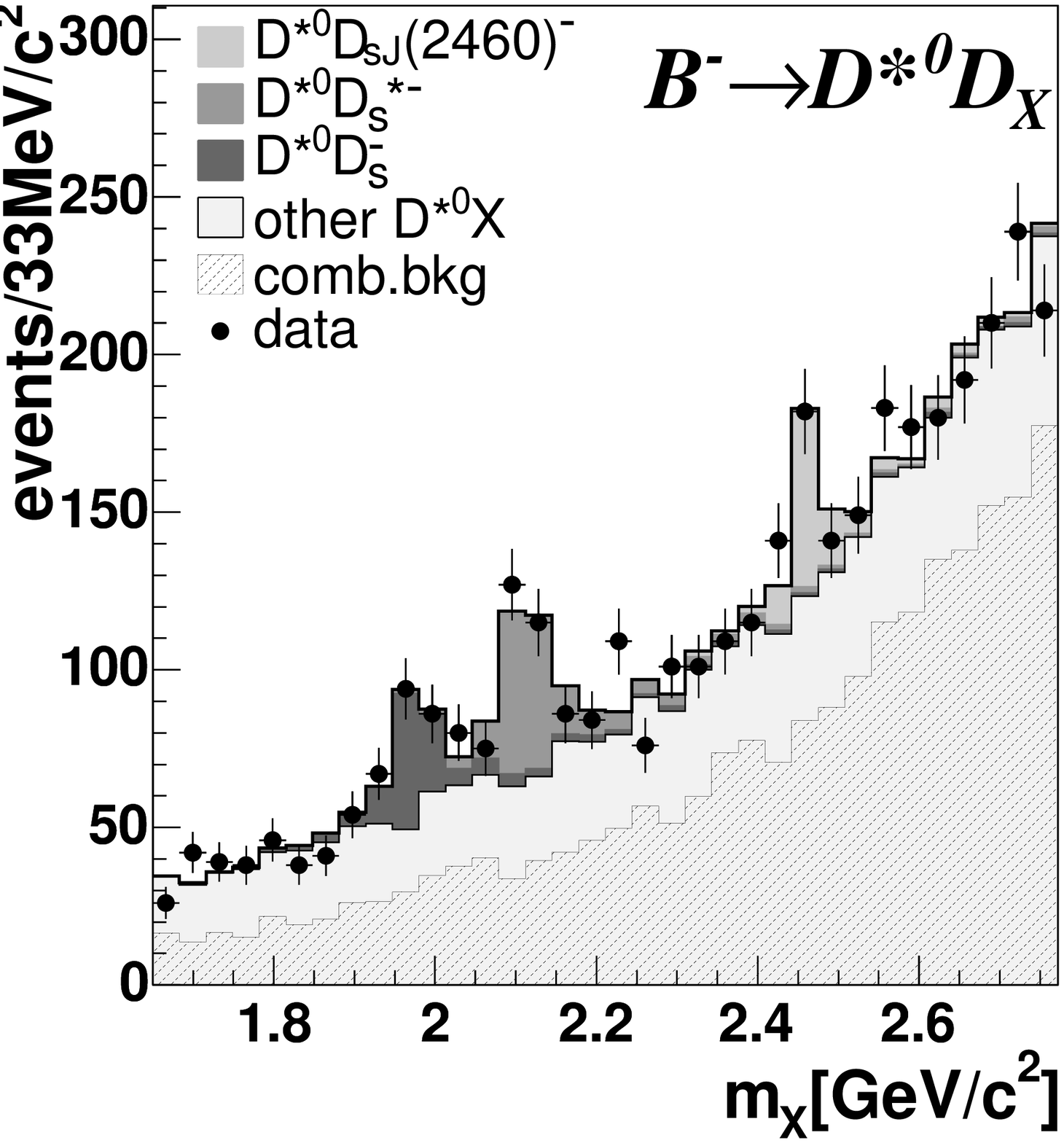}
\includegraphics[width=0.49\linewidth]{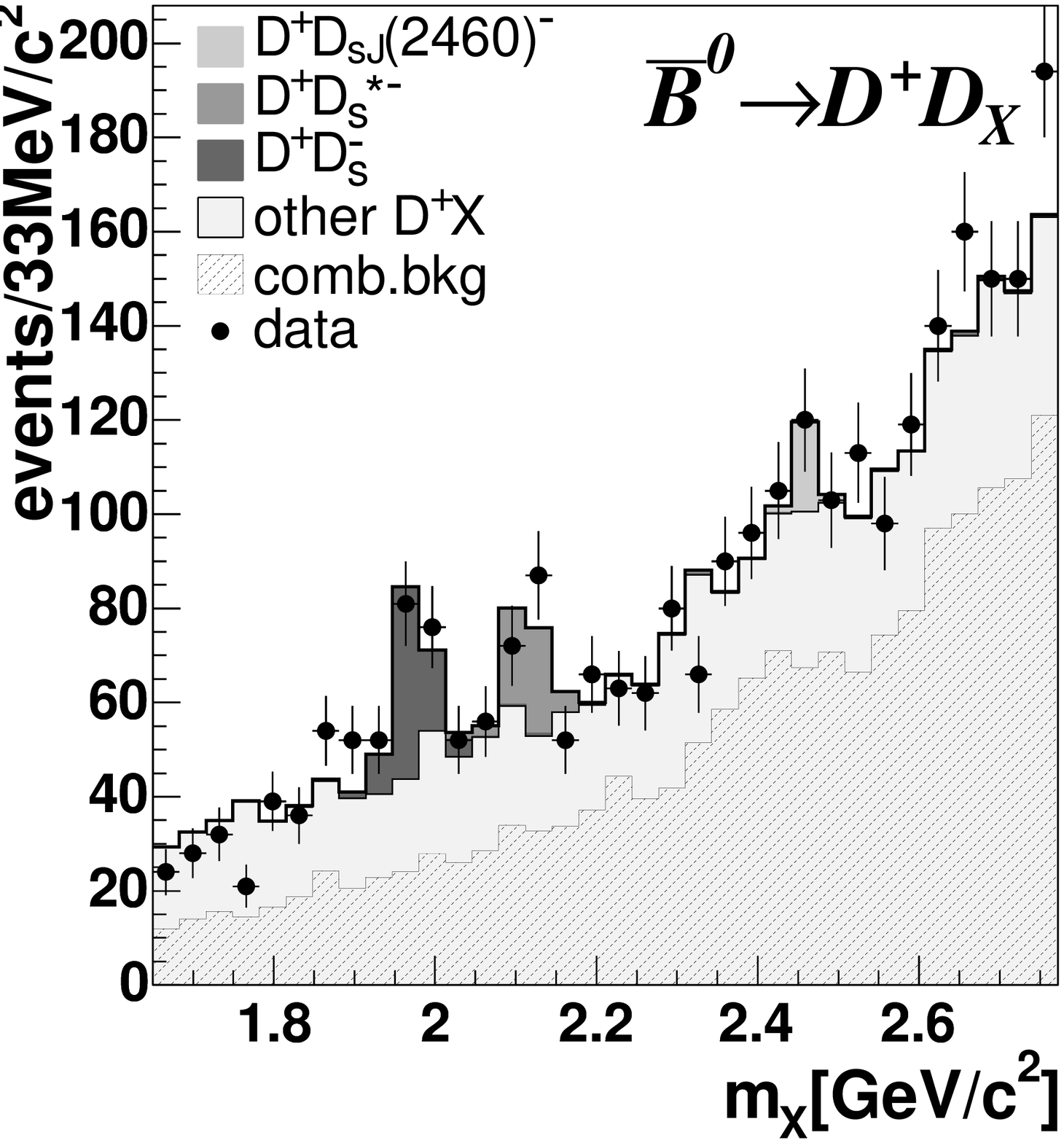}
\includegraphics[width=0.49\linewidth]{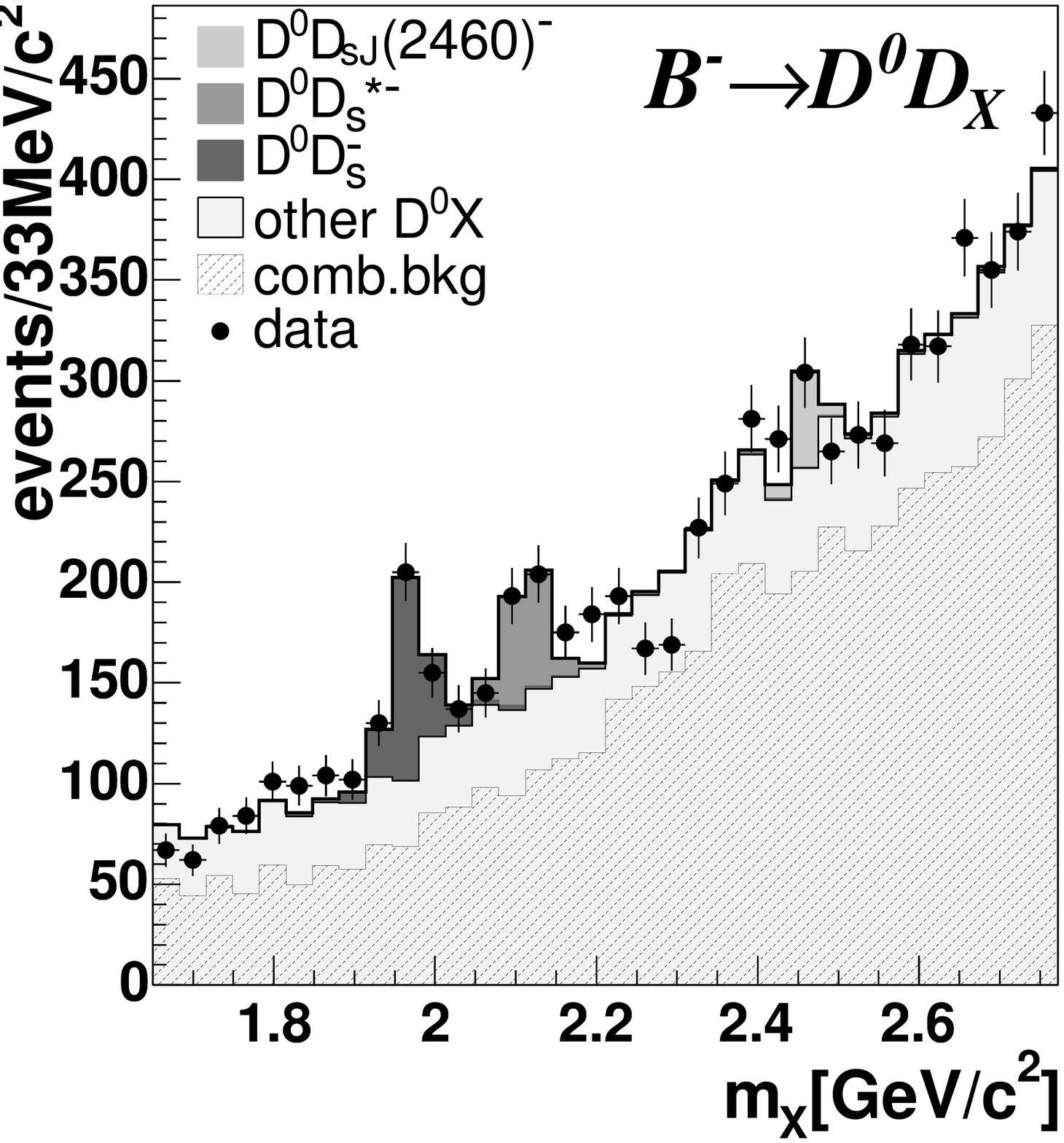}

\end{minipage}
\caption{The recoil mass against a $D$ or  $D^*$. 
(From Ref.~\cite{babar_ds_new}.)}
\label{fig:babar_d_recoil}
\end{center}
\end{figure}

\begin{figure}[tb]
\begin{center}
\begin{minipage}{\linewidth }
\includegraphics[width=0.49\linewidth]{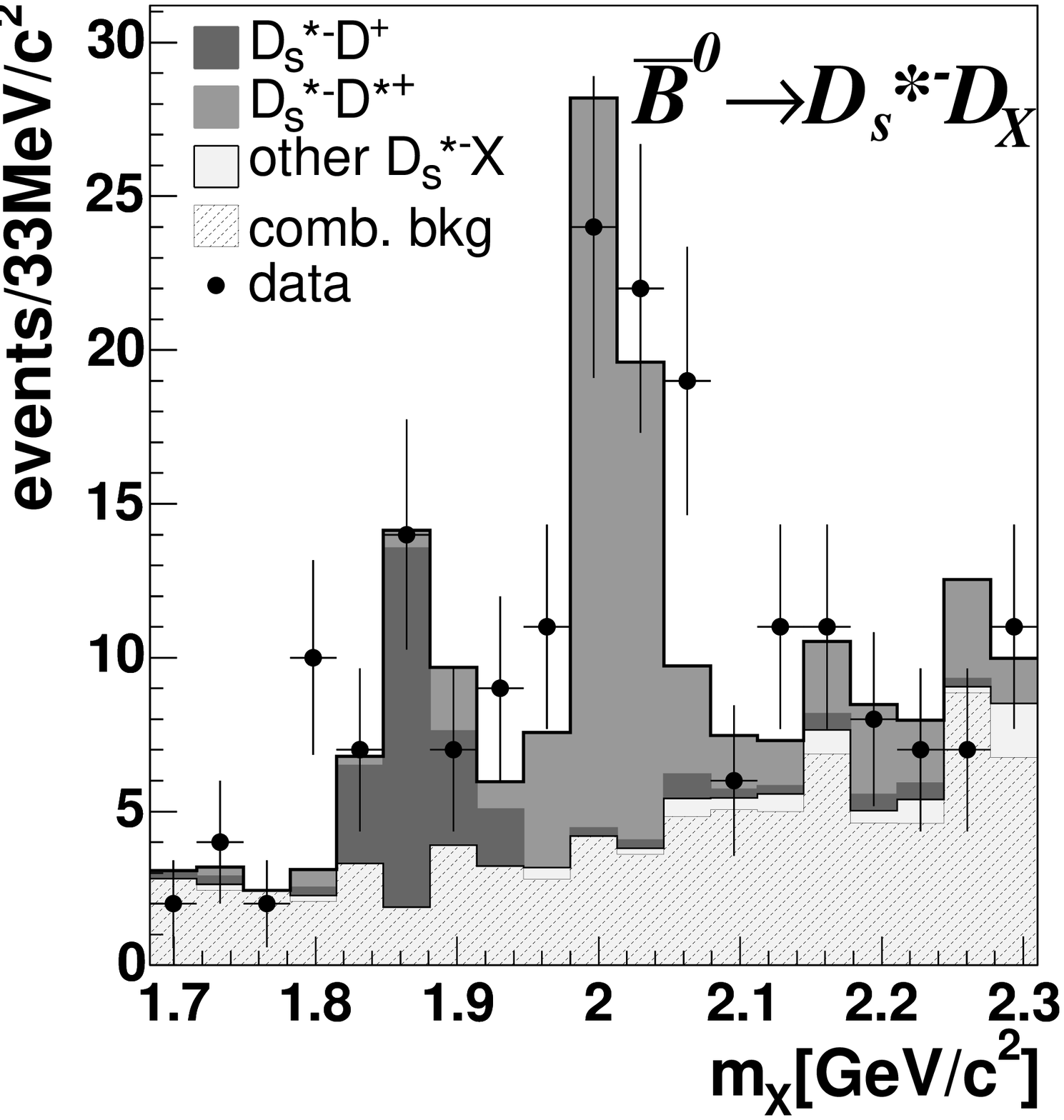}
\includegraphics[width=0.49\linewidth]{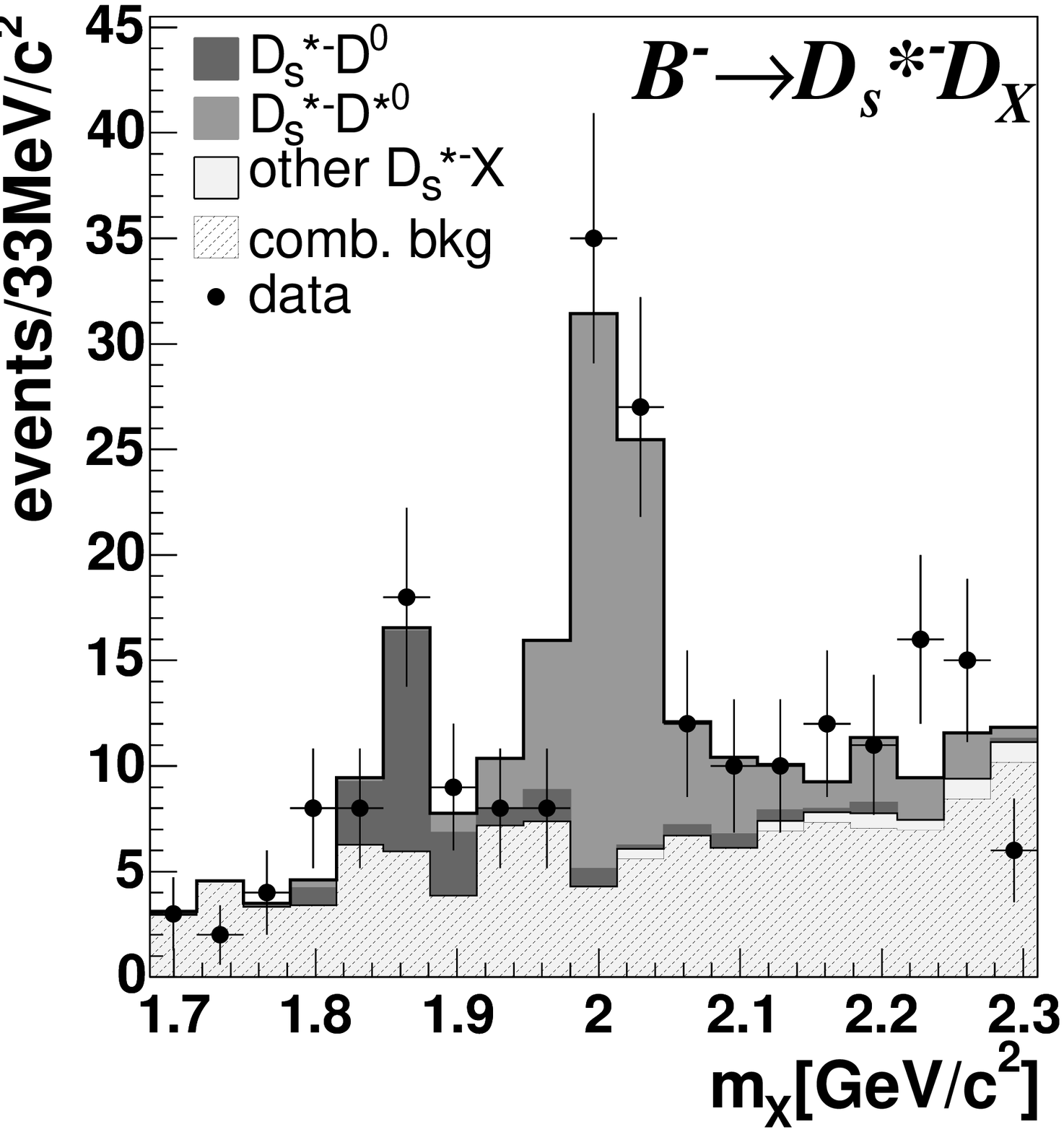}
\includegraphics[width=0.49\linewidth]{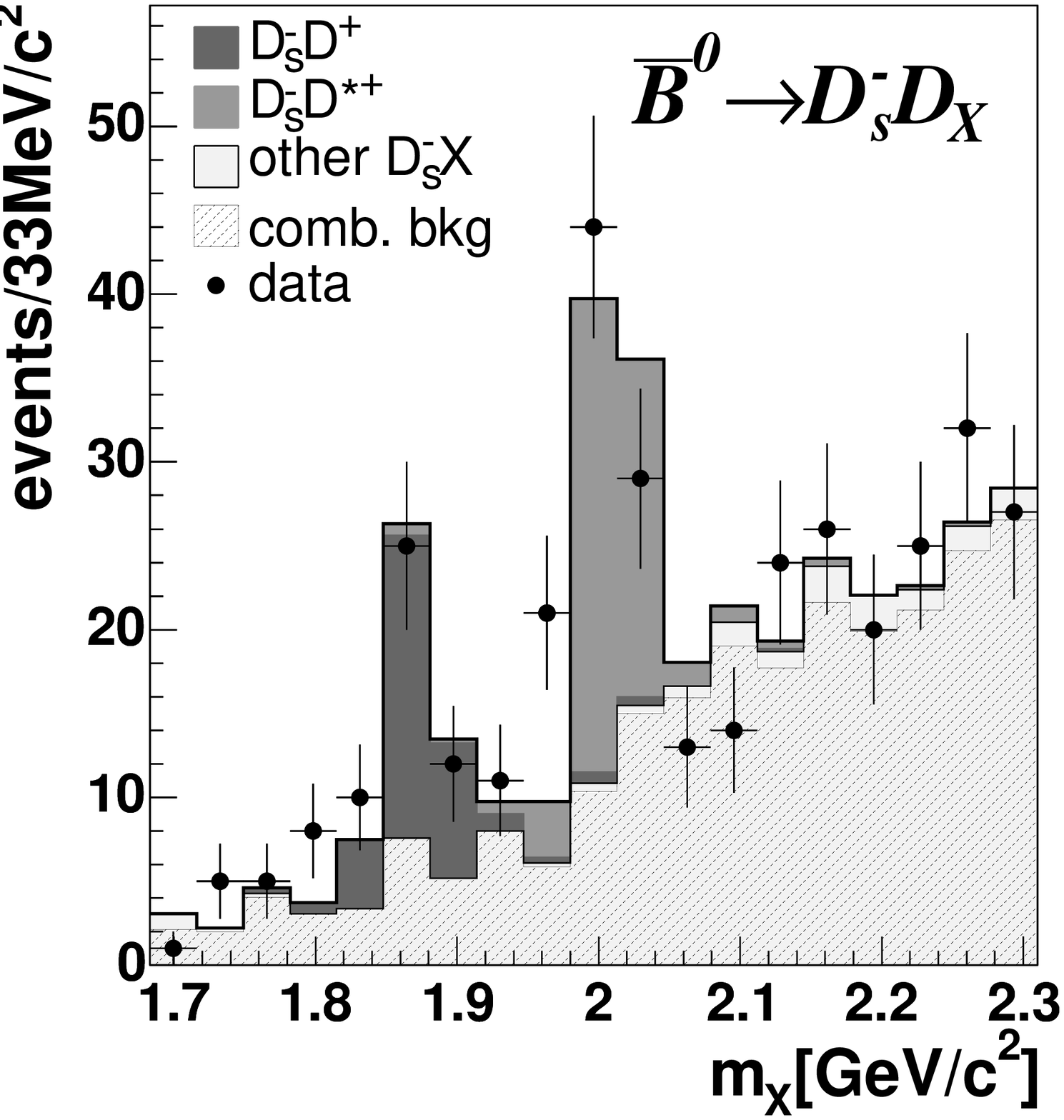}
\includegraphics[width=0.49\linewidth]{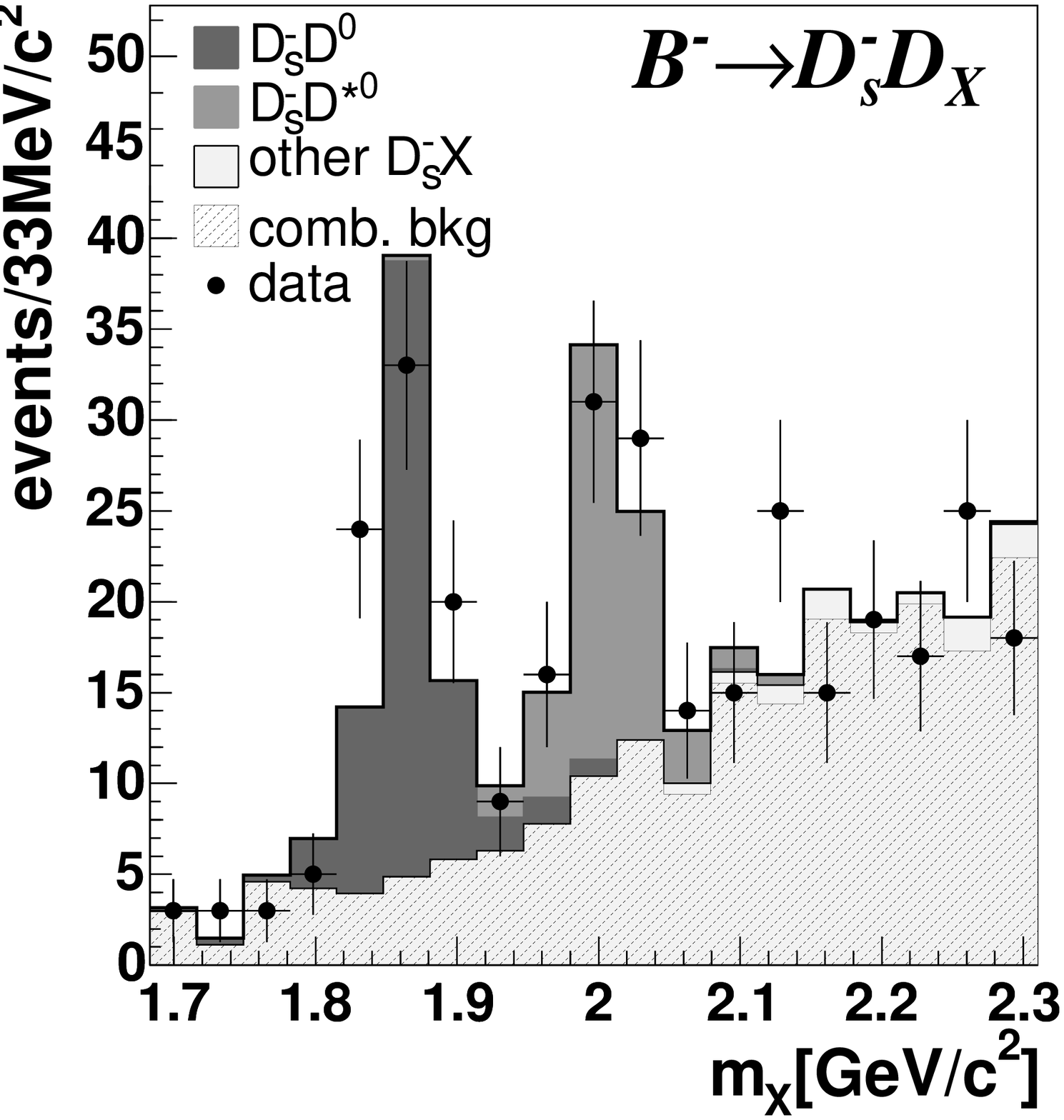}
\end{minipage}
\caption{The recoil mass against a $D_s$ or  $D_s^*$
(From Ref.~\cite{babar_ds_new}.)}
\label{fig:babar_ds_recoil}
\end{center}
\end{figure}

An earlier BABAR study has used $B\to D^*D_s^*$ decays and
a technique of partially reconstructing either the $D^*$ or the
$D_s^*$ to measure the $D_s\to \phi\pi$ branching fraction\cite{babar_dsphipi}.
They quote ${\cal B}(D^+_s\to\phi\pi^+)=(4.81\pm0.52\pm0.38)\%$
based on a sample of $123\times 10^6$ $B\bar B$ decays. 
More recently BABAR\cite{babar_ds_new} has
presented preliminary results based on 210 fb$^{-1}$ of
data where they use a tag technique in which one $B$ is 
fully reconstructed. In events with one fully reconstructed
$B$ candidate BABAR reconstructs one additional $D^{(*)}$ or
$D_{s(J)}^{(*)}$ meson. Then they look at the recoil mass against
this reconstructed candidate. The recoil masses are shown
in Figs.~\ref{fig:babar_d_recoil} and~\ref{fig:babar_ds_recoil}.

From these modes BABAR extracts ${\cal B}(D_{sJ}(2460)^-\to D_s^{*-}\pi^0)
=(56\pm13\pm9)\%$ and ${\cal B}(D_{sJ}(2460)^-\to D_s^{*-}\gamma)
=(16\pm4\pm3)\%$ in addition to ${\cal B}(D_s^-\to \phi\pi^+)=
(4.62\pm0.36\pm0.50)\%$.

\section{Inclusive measurements of $\eta$, $\eta'$, and $\phi$ production in $D$ and $D_s$ decays}

Using samples of tagged $D$ and $D_s$ decays CLEO-c has measured
the inclusive production of  $\eta$, $\eta'$, and $\phi$ mesons
by looking at the recoil against the tag\cite{cleoc_inclusive}. The results are
summarized in Table~\ref{tab:cleoc_inclusive}. The knowledge of
inclusive measurements before this CLEO-c measurement was
poor, besides limits only ${\cal B}(D^0\to \phi X)=1.7\pm0.8$ 
was measured. As expected the $\eta$, $\eta'$, and $\phi$ rates
are much higher in $D_s$ decays.

\begin{table}[bt]
\caption{Inclusive branching fractions}
\label{tab:cleoc_inclusive}
\begin{center}
\begin{tabular}{lc}
\hline\hline
Decay  & ${\cal B}$ (\%) \\ \hline
$D^0\to\eta X$          & $9.5\pm0.4\pm0.8$        \\
$D^-\to\eta X$          & $6.3\pm0.5\pm0.5$         \\
$D_s^+\to\eta X$        & $23.5\pm3.1\pm2.0$         \\
\hline
$D^0\to\eta' X$         & $2.48\pm0.17\pm0.21$        \\
$D^-\to\eta' X$         & $1.04\pm0.16\pm0.09$         \\
$D_s^+\to\eta' X$       & $8.7\pm1.9\pm1.1$         \\
\hline
$D^0\to\phi X$          & $1.05\pm0.08\pm0.07$        \\
$D^-\to\phi X$          & $1.03\pm0.10\pm0.07$         \\
$D_s^+\to\phi X$        & $16.1\pm1.2\pm1.1$         \\
\hline\hline
\end{tabular}
\end{center}
\end{table}

\section{The doubly Cabibbo suppressed decay $D^+\to K^+\pi^0$}

Both CLEO-c and BABAR have studied the doubly Cabibbo suppressed
decay $D^+\to K^+\pi^0$. CLEO-c\cite{cleoc_dcsd} has reconstructed candidates in a
281 pb$^{-1}$ sample of $e^+e^-$ data recorded at the $\psi(3770)$. 
BABAR\cite{babar_dcsd}
has used a sample of 124 fb$^{-1}$ recorded at the $\Upsilon(4S)$.
CLEO-c and BABAR finds branching fractions in good agreement with
each other, 
${\cal B}(D^+\to K^+\pi^0)=(2.24\pm0.36\pm0.15\pm0.08)\times 10^{-4}$
and ${\cal B}(D^+\to K^+\pi^0)=(2.52\pm0.46\pm0.24\pm0.08)\times 10^{-4}$
respectively. 

\section{Modes with $K^0_L$ or $K^0_S$ in the final states}

It has commonly been assumed that $\Gamma(D\to K^0_S X)=\Gamma(D\to K^0_L X)$.
However, as pointed out by Bigi and Yamamoto\cite{bigi} this is
not generally true as for many $D$ decays there are contributions from
Cabibbo favored and Cabibbo suppressed decays that interfere and
contributes differently to final states with $K^0_S$ and $K^0_L$.
As an example consider $D^0\to K^0_{S,L}\pi^0$. Contributions to
these final states involve the Cabibbo favored decay $D^0\to \bar K^0\pi^0$
as well as the Cabibbo suppressed decay $D^0\to K^0\pi^0$. However,
we don't observe the $K^0$ and the $\bar K^0$ but rather the $K^0_S$
and the $K^0_L$. As these two amplitudes interfere constructively to
form the $K^0_S$ final state we will see a rate asymmetry. Based
on factorization Bigi and Yamamoto predicted
\begin{eqnarray*}
R(D^0)&\equiv&{{\Gamma(D^0\to K^0_S\pi^0)-\Gamma(D^0\to K^0_L\pi^0)}\over
{\Gamma(D^0\to K^0_S\pi^0)+\Gamma(D^0\to K^0_L\pi^0)}}\\
     &\approx& 2\tan^2\theta_C\approx 0.11.\\
\end{eqnarray*}
Using tagged $D$ mesons CLEO-c has measured this asymmetry and obtained
$$
R(D^0)=0.122\pm0.024\pm0.030
$$
which is in good agreement with the prediction.

Similarly, CLEO-c has also measured the corresponding asymmetry in charged
$D$ mesons and obtained
\begin{eqnarray*}
R(D^+)&\equiv&{{\Gamma(D^+\to K^0_S\pi^+)-\Gamma(D^+\to K^0_L\pi^+)}\over
{\Gamma(D^+\to K^0_S\pi^+)+\Gamma(D^+\to K^0_L\pi^+)}}\\
   &=&0.030\pm0.023\pm0.025.\\
\end{eqnarray*}
Prediction of the asymmetry in charged $D$ decays is more 
involved. D.-N.~Gao predicts~\cite{gao} this asymmetry to be in the 
range 0.035 to 0.044, which is consistent with the observed asymmetry.

\section{Summary}

 Recently there has been a lot of progress on the determination
of absolute hadronic branching fractions of $D$ and $D_s$ mesons.
Here recent results from CLEO-c and the B-factory experiments, BABAR
and Belle, 
were reported. CLEO-c uses the extremely clean environment 
at threshold for these measurements while the B-factory
experiments use their very large data samples to explore 
partial reconstruction techniques to determine the absolute 
hadronic branching fractions.

\section*{Acknowledgments}
This work was supported by the National Science Foundation grant
PHY-0202078 and by the Alfred P.~Sloan foundation.

\bigskip 

\end{document}